\documentclass[twocolumn]{aastex631}

\shorttitle{OverCite}
\shortauthors{OverCite}

\begin{document}

\title{\Large {\tt OverCite}: Add citations in \LaTeX\ without leaving the editor}

\author[0000-0003-1247-9349]{Cheyanne Shariat}
\affiliation{Department of Astronomy, California Institute of Technology, 1200 East California Boulevard, Pasadena, CA 91125, USA}
\email{cshariat@caltech.edu}

\begin{abstract}
Adding citations while drafting in \LaTeX\ often requires leaving the editor, searching for a paper in mind, copying its BibTeX entry into the project bibliography, renaming the cite key, and then returning to the sentence. \texttt{OverCite} is an open-source, lightweight tool that lets authors find, select, and insert citations without leaving the writing environment. 
In Overleaf, \texttt{OverCite} uses rough citation placeholders (e.g., $\texttt{\textbackslash citep\{Perlmutter1999\}}$) and local sentence context to query ADS/SciX-indexed literature, rank likely matches, and insert the selected reference, without leaving the editor. A companion \texttt{VS Code} extension provides the same functionality for local \LaTeX\ projects. The ADS/SciX database includes astronomy, physics, computer science, mathematics, biology, and \emph{all} indexed arXiv e-prints, making \texttt{OverCite} useful across a broad range of scientific disciplines. 

\end{abstract}

\section{Introduction}

Citations are central to scientific writing.
In practice, drafting scientific writing frequently involves locating and inserting references.
For many \LaTeX\ users, that step interrupts the flow of writing. A sentence is written, a citation is needed, and the author leaves the editor to find a paper already in mind, export the BibTeX entry, and return to the manuscript. This interruption is particularly avoidable when the intended citation is already known, and the remaining task is to simply locate and insert the reference.

\texttt{OverCite} was developed to accelerate this aspect of scientific writing. 
\texttt{OverCite} is a citation tool designed for the scenario where an author already knows, at least roughly, what citation belongs in a sentence and wants to insert it with minimal interruption. It queries the ADS/SciX literature database, ranks likely matches, and inserts the corresponding BibTeX entry directly into the project bibliography, all without leaving the editor (Figure~\ref{fig:workflow}).

\section{OverCite Workflow}

\begin{figure*}[t]
\centering
\includegraphics[width=0.99\textwidth]{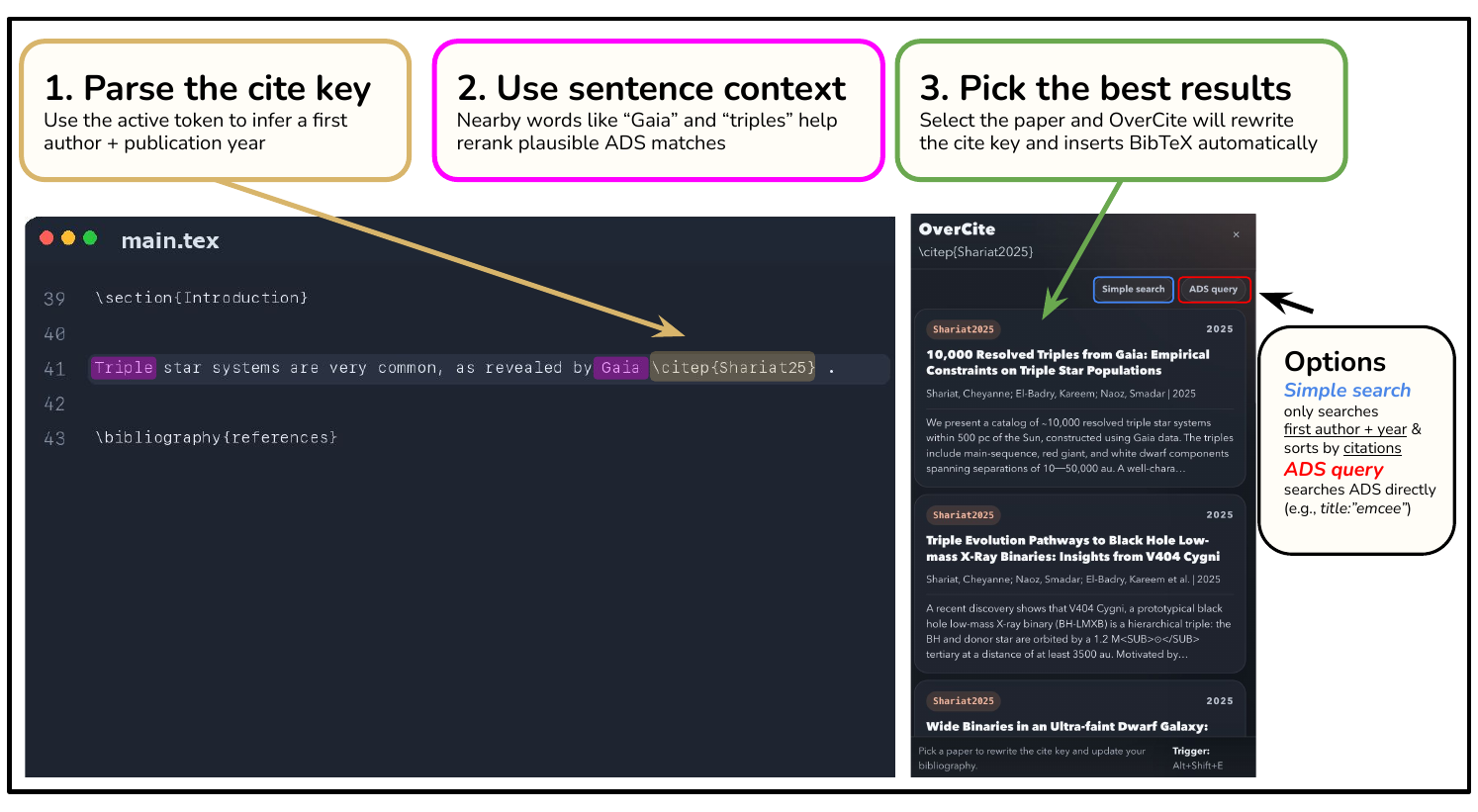}
\caption{
\textbf{\texttt{OverCite} workflow.} A partial citation key in the manuscript (yellow) is interpreted together with the surrounding sentence context (purple), and ADS/SciX matches are presented in a popup (right). The user can then select a record and insert its BibTeX entry directly into the project bibliography. \texttt{OverCite} supports three search modes: (1) the default contextual mode (shown here), (2) a simple-search mode (blue) that ignores sentence context and instead searches typed author/year information, and (3) an ADS-query mode (red) that sends the typed token directly to ADS/SciX. The intended citation for this particular example is the first option, \citet{Shariat2025}.
}
\label{fig:workflow}
\end{figure*}

\texttt{OverCite} is available both as an Overleaf browser extension and as a companion \texttt{VS Code} extension for local projects\footnote{The package is open source and publicly available on GitHub at \url{https://github.com/cheyanneshariat/OverCite}, with a frozen version archived on Zenodo \citep{Shariat_OverCite_2026}.}.
Figure~\ref{fig:workflow} illustrates the core workflow.
The user places the cursor inside a citation command and triggers \texttt{OverCite} from within the editor. The active citation token is then parsed and used to search ADS/SciX. \texttt{OverCite} currently supports three search modes: 
\begin{enumerate}
    \item \textbf{contextual search}:~uses the typed token together with the local sentence
    \item \textbf{simple search}:~ignores sentence context and searches typed author/year information alone (sorted by citation count)
    \item \textbf{ADS-query}:~sends the typed token directly to ADS/SciX
\end{enumerate}
Candidate papers are displayed in a popup, where the user can inspect the results and select the intended paper. Once a paper is chosen, \texttt{OverCite} updates the citation key in the manuscript and inserts the corresponding BibTeX entry into the project bibliography.

The citation-insertion process is adaptable to different drafting habits and project setups. The settings also let users choose a preferred citation-key style, control bibliography entry order, and specify the target bibliography file in projects with multiple \texttt{.bib} files. These options allow the same basic workflow to support a range of \LaTeX\ citing habits.

The public implementation of \texttt{OverCite} is deterministic and does not use a large language model. It only uses typed citation cues, local sentence context, and rule-based ranking of ADS/SciX results (for the full ranking logic, see the public repository).

\subsection{Representative Use Cases}

A common use case is a direct author-year lookup, for example \texttt{\textbackslash citep\{Shariat25\}} (shown in Figure \ref{fig:workflow}), where the citation token already contains an author and publication year. In that scenario, the surrounding sentence mainly serves as an additional check when ranking likely matches. Another common use case is disambiguation of common surnames, such as \texttt{\textbackslash citep\{Smith25\}}, where many plausible papers may share the same author-year combination. In those cases, the surrounding sentence usually provides the additional information needed to favor the intended paper. However, when further filtering is needed, \texttt{OverCite} also allows a first initial in the citation key, for example \texttt{\textbackslash citep\{SmithJ25\}}.

\texttt{OverCite} also supports less informative citation keys. This includes omitting the publication year (e.g.,~\texttt{\textbackslash citep\{Abbott\}}), including full year (e.g.,~\texttt{\textbackslash citep\{Hawking1975\}}), and surnames of non–first authors can also be searched, although such matches generally rank below first-author matches. It also applies for collaborations, such as \texttt{\textbackslash citep\{Astropy Collaboration\}} or \texttt{\textbackslash citep\{Gaia Collaboration2021\}}.
These patterns allow \texttt{OverCite} to accommodate a range of drafting habits while remaining centered on the same basic task: identifying and inserting the correct citation.

For cases where the intended ADS/SciX query requires a direct search, \texttt{OverCite} also supports an ADS query mode. This allows the citation token itself to act as a literal ADS/SciX search string, for example \texttt{\textbackslash citep\{title:"emcee"\}} or \texttt{\textbackslash citep\{author:"schlegel" maps of dust\}}. This mode is useful when the author wants finer control over the search while still staying inside the editor.

\section{Scope and Outlook}
\texttt{OverCite} is designed for citation insertion during drafting, especially when an author already knows which reference they want to cite.
Reference managers such as Zotero and Mendeley are designed to collect, organize, and maintain bibliographies. \texttt{OverCite}, by contrast, focuses on the in-editor step of locating the paper, inserting its BibTeX entry into the project bibliography, and updating the citation key. It is therefore best suited for querying known papers during writing and is explicitly \textit{not} intended to replace the broader process of paper exploration, literature review, or discovery of unfamiliar work.

The package is open source and publicly available at \url{https://github.com/cheyanneshariat/OverCite}. Its current public implementation (\texttt{v0.1.3}) supports use in Overleaf and in local \LaTeX\ environments through \texttt{VS Code} (see Section \ref{availability}).
Its dependence on ADS/SciX makes it immediately useful across a broad range of scientific disciplines, including all arXiv e-prints \citep[][]{ads_about, scix_about}. Each user authenticates with their own ADS/SciX token, so API usage is distributed per user rather than routed through a shared \texttt{OverCite} backend. Currently, ADS allows $5000$ regular requests per day for most request types, with higher limits available by request \citep{ads_rate_limits}. In future versions, the same framework could be extended to support additional editors/browsers, use in other writing environments beyond \LaTeX, alternative bibliography styles, refined search and ranking logic, and additional literature databases where useful.

\section{Software Availability} \label{availability}
\texttt{OverCite} is available both as an Overleaf browser extension and as a companion \texttt{VS Code} extension for local projects.
It is available as a \href{https://chromewebstore.google.com/detail/overcite/hmjojciemhnfkjnilakhehkgkhkplbdo}{Chrome extension}, 
\href{https://addons.mozilla.org/en-US/firefox/addon/overcite/?utm_source=addons.mozilla.org&utm_medium=referral&utm_content=search}{Firefox Add-on}, or \href{https://marketplace.visualstudio.com/items?itemName=CheyanneShariat.overcite-vscode}{\texttt{VS Code} extension}, with a frozen version on Zenodo \citep{Shariat_OverCite_2026}.

\section{Acknowledgments} \label{acknowledgments}
I thank Kareem El-Badry for useful feedback on the manuscript as well as Isaac Cheng, Sam Whitebook, and Cameron Hummels for inspiring discussions and comments on early versions of the tool.
I acknowledge support from the Department of Energy Computational Science Graduate Fellowship under Award Number DE-SC0026073, supported by the U.S. Department of Energy, Office of Science, and Office of Advanced Scientific Computing Research. 

\bibliographystyle{aasjournal}
\bibliography{refs}

@misc{ads_about,
  author = {{SAO/NASA Astrophysics Data System}},
  title = {About {ADS}},
  year = {2026},
  howpublished = {\url{https://ui.adsabs.harvard.edu/about/}},
  note = {Accessed: 2026-03-19}
}

@misc{scix_about,
  author = {{SciX}},
  title = {SciX: Integrated Access to Research Across Earth, Environmental, and Space Sciences},
  year = {2025},
  howpublished = {\url{https://scixplorer.org/help/whats_new/}},
  note = {Accessed: 2026-03-19}
}

@ARTICLE{Shariat2025,
       author = {{Shariat}, Cheyanne and {El-Badry}, Kareem and {Naoz}, Smadar},
        title = "{10,000 Resolved Triples from Gaia: Empirical Constraints on Triple Star Populations}",
      journal = {\pasp},
         year = 2025,
        month = sep,
       volume = {137},
       number = {9},
          eid = {094201},
        pages = {094201},
          doi = {10.1088/1538-3873/adfb30},
archivePrefix = {arXiv},
       eprint = {2506.16513},
 primaryClass = {astro-ph.SR},
       adsurl = {https://ui.adsabs.harvard.edu/abs/2025PASP..137i4201S}
}

@misc{ads_rate_limits,
  author       = {{NASA Astrophysics Data System}},
  title        = {API Rate Limit Policy},
  year         = {2026},
  howpublished = {\url{https://ui.adsabs.harvard.edu/help/policies/rate-limits}},
  note         = {Accessed 2026-03-24}
}

@software{Shariat_OverCite_2026,
  author   = {Shariat, Cheyanne},
  title    = {OverCite: Add citations in LaTeX without leaving the editor},
  year     = {2026},
  version  = {0.1.5},
  doi      = {10.5281/zenodo.19485160},
  url      = {https://zenodo.org/records/19485160},
  publisher = {Zenodo}
}

\end{document}